\documentclass[prd, preprint, longbibliography, 12pt]{revtex4-1}

\usepackage{amsmath}
\usepackage{mathtools}
\usepackage{amssymb}
\usepackage{setspace}
\usepackage{graphicx}
\usepackage{natbib}
\usepackage{float}
\usepackage{tensor}
\usepackage[utf8]{inputenc}
\usepackage{amsfonts}
\usepackage{braket}
\usepackage{esint}
\usepackage{breqn}
\usepackage{IEEEtrantools}

\begin{document}
 
 % ! TEX spellcheck
 %
%\ \vskip 1.0 in

\begin{center}
{ \large \bf The exceptional Jordan algebra, and its implications for our understanding of gravitation and the weak force}

\smallskip

\vskip 0.1 in

{\large{\bf Tejinder P.  Singh$^{1,2}$ }}

\smallskip

${}^1${\it Inter-University Centre for Astronomy and Astrophysics,}\\ {\it Post Bag 4, Ganeshkhind,  Pune 411007, India}\\
${}^2${\it Tata Institute of Fundamental Research,}\\
{\it Homi Bhabha Road, Mumbai 400005, India}\\
\smallskip
 {\tt Email:  tejinder.singh@iucaa.in, tpsingh@tifr.res.in}

\end{center}
\vskip 1 in

\centerline{\bf ABSTRACT}
\smallskip

\noindent The exceptional Jordan algebra is the algebra of $3 \times 3$ Hermitian matrices with octonionic entries. It is the only one from Jordan’s algebraic formulation of quantum mechanics which is not equivalent to the conventional formulation of quantum theory. It has often been suggested that this exceptional algebra could explain physical phenomena not currently explained by the conventional approach, such as values of the fundamental constants of the standard model of particle physics, and their relation to gravitation. We show that this is indeed the case; and this also unravels the connection between general relativity and the weak force. The exceptional Jordan algebra also predicts a new U(1) gravitational interaction which modifies general relativity, and which provides a theoretical basis for understanding the Modified Newtonian Dynamics (MOND). 

\noindent 
\vskip 1 in

\centerline{March 28, 2023}

\bigskip

\centerline{Essay written for the Gravity Research Foundation 2023 Awards for Essays on Gravitation}
%\centerline{This essay received an Honorable Mention}

\newpage

\bigskip

\noindent In the year 1934, Pascual Jordan, John von Neumann and Eugene Wigner published their celebrated paper titled ``On an Algebraic Generalisation of the Quantum Mechanical Formalism" \cite{jnw}. Building on earlier work by Jordan, they were investigating an algebra of observables defined by the Jordan product
\begin{equation}
A \circ B \equiv \frac{1}{2} \left[ A \times B + B \times A \right]
\end{equation}
When $A$ and $B$ are Hermitian matrices, their Jordan product is necessarily Hermitian, unlike their matrix product $A \times B$ which need not be Hermitian. Under the Jordan product, Hermitian matrices form a closed algebra which is commutative, but non-associative, though power associative. The property of power associativity ensures that if some observable $A$ takes the value $a$ then a function $f(A)$ of $A$ takes the value $f(a)$. These physically desirable features make the Jordan algebraic formulation of quantum mechanics an attractive alternative formulation.

The Jordan product has the property 
\begin{equation}
A \circ (A^2 \circ B) = A^2 \circ (A \circ B)
\end{equation}
where $A^2 = A \circ A$ is the same as the matrix product of $A$ with itself. If $A$ and $B$ are matrices, this relation is an identity, known as the Jordan identity. Nonetheless, the Jordan algebra can be directly defined as a vector space with a commutative bi-linear product obeying the Jordan identity. And after defining a multiplicative identity element, it serves as an algebra of observables.

In the aforesaid paper, Jordan et al. showed that this algebra is equivalent to the ordinary matrix algebra of Hermitian matrices (and hence equivalent to conventional quantum mechanics) of arbitrary dimension, with entries as real numbers, complex numbers, or quaternions. However, there is one exception to this equivalence, that being the Jordan algebra of $3\times 3$ Hermitian matrices with octonions as entries. This is known as the exceptional Jordan algebra (also called the Albert algebra) and denoted $J_3(8)$. The authors conclude that `essentially new results, in contrast to the present content of quantum mechanics' are only to be expected in this one exceptional case. Several decades later, Townsend \cite{Tsend} airs the same sentiment: ``The octonionic quantum mechanics based on the exceptional Jordan algebra \cite{Gunay} is the most interesting case because
exceptionality implies that no Hilbert space formulation is possible. This is therefore
a genuine, and radical, generalisation of quantum mechanics, although one without as
yet any application."

In this essay we show that the exceptional case indeed has an application, one of far-reaching consequences, which impacts on how we understand space-time geometry and gravitation, as well as the weak force. To get there, we begin by recalling the following insightful remark from Edward Witten \cite{Witten}:

\smallskip

\noindent ``If one wants to summarise our knowledge of physics in the briefest possible terms, there are three really fundamental observations: (i) Space-time is a pseudo-Riemannian manifold $M$, endowed with a metric tensor and governed by geometrical laws. (ii) Over $M$ is a vector bundle $X$ with a nonabelian gauge group $G$. (iii) Fermions are sections of $(\hat{S}_{+} \otimes V_R)\oplus(\hat{S}_{-}\otimes V_{\bar{R}})$. $R$ and $\bar{R}$ are not isomorphic; their failure to be isomorphic explains why the light fermions are light and presumably has its origins in a representation difference $\Delta$ in some underlying theory. All of this must be supplemented with the understanding that the geometrical laws obeyed by the metric tensor, the gauge fields, and the fermions are to be interpreted in quantum mechanical terms''  [{\it from the CERN preprint `Physics and Geometry' (1987)}]. 

It turns out that Witten's sought for `some underlying theory' is  based precisely on the algebra of the octonions, and fermions must be interpreted not in `quantum mechanical terms' but in terms of its aforesaid radical generalisation, the exceptional Jordan algebra! This also leads to a unification of the space-time manifold $M$ and the vector bundle $X$ with gauge group $G$ into a larger bioctonionic space possessing $E_8 \times E_8$ symmetry.

These conclusions originate from the following quantum foundational considerations. The time parameter employed to describe evolution in quantum theory is part of a classical space-time. The latter exists if and only if the universe is dominated by macroscopic bodies obeying the laws of classical physics. For, if every elementary particle in the universe was to be in a quantum superposition of two or more position states, this can be expected to result in quantum gravitational states which are necessarily  a superposition of classical gravitational states. The point structure of the underlying space-time manifold is then lost: because operational distinguishability of space-time points requires the manifold to be overlaid by a classical gravitational field (the Einstein hole argument). This argument is valid at all energy scales, not just at the Planck energy scale. Our low energy universe happens to be dominantly classical, but in principle it need not be so. Therefore, even at the energies accessible in current experiments, there must exist a reformulation of quantum mechanics which does not depend on classical time. In our investigations, we arrive at more than a reformulation; in fact at the `radical generalisation' afforded by the exceptional Jordan algebra. This yields significantly more information about the standard model of particle physics, and about gravitation, compared to the information  currently available though relativistic quantum field theory, and through the theory of general relativity. 

The sought for exceptional generalisation of quantum dynamics has only three independent fundamental constants: Planck length $L_p$, Planck time $\tau_p$, and Planck's constant $\hbar$. Compared to conventional approaches to quantum gravity, we have traded Planck mass $m_p$ for $\hbar$. And for good reason. If all physical subsystems in the universe have an action of the order $\hbar$ each, then space-time cannot have its classical point structure, energy scale notwithstanding. If the length and time scales of interest are also Planck scale, such a universe will be Planck energy scale (and hence quantum gravitational), but not if the length / time scales are much larger than Planck length / time. The universe can lose its point structure even without being quantum gravitational. Only when the cosmos is dominated by classical bodies such as stars and galaxies, does the point structure of the classical spacetime manifold emerge at low energies. And classicality is determined by the degree of entanglement, not by energy scale.

The search for a replacement of the space-time manifold in quantum theory is guided by the principles of non-commutative geometry, and by the fact that there are only four division algebras. These are the reals $\mathbb R$, the complex numbers $\mathbb C$, the quaternions $\mathbb H$ and the octonions $\mathbb O$. The algebra of quaternions and of octonions, being non-commutative, affords the possibility that using these number systems as coordinates (instead of the reals) could help arrive at `quantum theory without classical time'. Great help comes from Connes' non-commutative geometry, where it has been shown that the Tomita-Takesaki theorem applied to von Neumann algebras permits the existence of a one-parameter family of algebra automorphisms which play the role of time. This time, which we call Connes time and denote $\tau$, is a unique feature of non-commutative geometries, a feature that is absent in Riemannian geometries \cite{Connes}. The importance of (complex) octonions has also emerged from their successful use in the definitions of quarks and leptons of the standard model \cite{Furey1, Furey2}: the symmetries of the octonion algebra possibly contain within themselves the symmetries of the standard model.

With this motivation, we propose to use the octonions as coordinates of a non-commutative manifold (on which gauge fields and gravitation and elementary particles live). We define a 16D vector space by using the split bioctonion:
\begin{equation}
\begin{split}
\mathbb{O} \oplus \omega \mathbb{\tilde O}= & (a_0 + a_1 {\bf e}_1 + a_2 {\bf e}_2 + a_3 {\bf e}_3 + a_4 {\bf e}_4 + a_5 {\bf e}_5 + a_6 {\bf e}_6 + a_7 {\bf e}_7) \ \oplus \\
& \omega(a_0 - a_1 {\bf e}_1 - a_2 {\bf e}_2 - a_3 {\bf e}_3 - a_4 {\bf e}_4 - a_5 {\bf e}_5 - a_6 {\bf e}_6 - a_7 {\bf e}_7)
\end{split}
\label{spbioc}
\end{equation}
where $\omega$ is the split complex number, made here using the imaginary directions of the octonion. Every `atom of space-time-matter' carries such a coordinate system with itself, where an STM atom is an elementary (fermionic) particle such as an electron, along with all the bosonic fields that it produces. Dynamics is described in terms of matrix-valued coefficients of the bioctonion directions. Every dynamical variable is a matrix (with complex Grassmann numbers as its entries) which is a sum of a bosonic part and a fermionic part, and each of these two parts has sixteen matrix-valued components, one per each of the sixteen directions. For instance
\begin{equation}
Q_B = Q_0 + Q_1 {\bf e}_1 + Q_2 {\bf e}_2 + Q_3 {\bf e}_3 + Q_4 {\bf e}_4 + Q_5 {\bf e}_5 + Q_6 {\bf e}_6 + Q_7 {\bf e}_7
\end{equation}
shows the matrix-valued components $Q_i$ of a bosonic matrix $Q_B$ over octonionic space.

Consider the modulus square of the split bioctonion:
\begin{equation}
\begin{split}
|O \oplus \omega \tilde{O}|^2 & = (\tilde O \ominus \omega O) \times (O \oplus  \omega \tilde O) = \tilde{O}O \oplus \omega \tilde O \tilde O \ominus  \omega O O \ominus O \tilde O   \\
& =  (a_0^2 + a_1^2 + a_2^2 + a_3^2 + a_4^2 + a_5^2 + a_6^2 + a_7^2)\  \oplus \\
&\ \ \ \   \omega (a_0^2 - a_1^2 - a_2^2 - a_3^2 - a_4^2 - a_5^2 - a_6^2 - a_7^2 +Im1) \ \ominus \\
& \ \ \ \ \omega (a_0^2 - a_1^2 - a_2^2 - a_3^2 - a_4^2 - a_5^2 - a_6^2 - a_7^2 + Im2)\  \ominus  \\
&\ \ \ \   \   (a_0^2 + a_1^2 + a_2^2 + a_3^2 + a_4^2 + a_5^2 + a_6^2 +a_7^2) 
\end{split}
\label{thirteen}
\end{equation}
Two of the four resulting quadratic forms are Euclidean, whereas the other two are Lorentzian with imaginary corrections. Their physical interpretation is arrived at as follows. To begin with, it is useful to recall the important group-theoretic relations: $SL(2, \mathbb C) \sim SO(1,3), \ SL(2,\mathbb H) \sim SO(1,5), \ SL(2, \mathbb O) \sim SO(1,9)$. Hence, space-times with dimensions 4, 6, 10 are especially significant. Also, the Clifford algebras $Cl(3)$ and $Cl(7)$ are special; they relate respectively to split biquaternions and 6D spacetime and chiral leptons; and to split bioctonions and 10D spacetime and chiral quarks/leptons \cite{Vatsalya}. The universe is a collection of STM atoms, each carrying its own split bioctonionic coordinate system.

The Lagrangian of an STM atom is the trace of a matrix polynomial made from the dynamical matrices, and is assumed to have an $E_8 \times E_8$ symmetry \cite{Priyank}. Each of the $E_8$ branches as follows, as a result of the quantum-to-classical transition which is also simultaneously responsible for the electro-weak symmetry breaking:
\begin{equation}
E_8 \longrightarrow SU(3) \times E_6 \quad {\rm interpreted\ as} \quad SU(3)_{space} \times E_6
\end{equation}
The two $SU(3)_{space}$, one from each of the $E_8$,  are mapped to the 16D split bioctonionic space of Eqn. (5) and are hence responsible for the origin of the non-commutative manifold after symmetry breaking (= quantum-to-classical transition = left-right symmetry breaking). A quantum system, irrespective of energy scale, obeys the unbroken $E_8 \times E_8$ symmetry, unless and until a measurement is performed on it by a classical measuring apparatus. 

The matter and gauge fields, including the gravitational field, arise from the two copies of $E_6$, and live on the split bioctonionic space. In fact, $E_6$ is the only one of the five exceptional Lie groups $G_2, F_4, E_6, E_7, E_8$ which has complex representations.  Each of the two $E_6$ branches as follows:
\begin{equation}
E_6 \rightarrow SU(3) \times SU(3) \times SU(3)
\end{equation}
The first $E_6$ is interpreted as follows, after a further branching (resulting from electroweak symmetry breaking) of the last of the three $SU(3)$s:
\begin{equation}
E_6 \rightarrow SU(3)_{genLH} \times SU(3)_{color} \times SU(2)_L \times U(1)_Y
\end{equation}
The second $E_6$ is interpreted as
\begin{equation}
E_6 \rightarrow SU(3)_{genRH} \times SU(3)_{grav} \times SU(2)_R \times U(1)_{Yg}
\end{equation}
The two $SU(3)_{gen}$ are for the three generations of quarks and leptons of the standard model \cite{Priyank}, one for left-handed particles and the other for right-handed particles. $SU(3)_{color}$ is for QCD, and $SU(3)_{grav}$ is a newly predicted QCD-like gravitational interaction. $SU(2)_L \times U(1)_Y$ is the electroweak sector broken down to the weak interaction and (unbroken) $U(1)_{em}$; $SU(2)_R$ is the precursor of general relativity, and $U(1)_{Yg}$ is the RH counterpart of $U(1)_Y$; the resulting unbroken $U(1)_{grav}$ is sourced by square-root of mass, and is the theoretical origin of Milgrom's MOND. The entire right-handed sector is dark and has exclusively gravitational interaction; it provides a left-right symmetric extension of the standard model, which includes general relativity. The representation theory corresponding to this branching is consistent with what we know of the standard model, and has been discussed in detail in \cite{Priyank}.

Spinors of the Clifford algebra $Cl(6)$ generated by complex octonions (while keeping one of the octonionic imaginaries fixed) define one generation of quarks and leptons of the standard model, under the unbroken symmetry $SU(3)_{color} \times U(1)_{em}$ \cite{Furey1}. If the left-handed neutrino is assumed Majorana, then these are left-handed fermions. Electric charge, defined as one-third of the eigenvalue of the $U(1)$ number operator (identified with $U(1)_{em}$) is shown to be quantised, as observed \cite{Furey2}. States for three generations of fermions are generated through $SU(3)_{gen}$ and as expected, electric charge for the second and third generations is quantised in the same way as for the first generation \cite{Singhfsc}.

If none of the octonionic imaginary directions is kept fixed, the Clifford algebra $Cl(7)$ is generated, and is related to complex split bioctonions \cite{Vatsalya}. $Cl(7)$ can be thought of as a direct sum of two copies of $Cl(6)$ which are parity inverses of each other. States from one of these two $Cl(6)$ have been interpreted in the previous paragraph. As for the second $Cl(6)$, its mathematical construction is identical, but interpretation is different. The new interpretation is motivated by the observation that the masses of the electron, up quark and down quark are in the ratio $1: 4: 9$, and hence the square-roots of their masses are in the ratio $\pm 1 : \pm 2 : \pm 3$, which is a flipping of their electric charge ratios $\pm 3 : \pm 2 : \pm 1$. This symmetry could not be a coincidence; hence for the right-handed quarks and leptons we interpret the $U(1)$ quantum number as square-root of mass $\pm \sqrt{m}$, with plus sign for right-handed particles, and minus sign for their left-handed antiparticles. The assignments of the RH electron and RH down quark are switched, compared to their LH counterparts, so that the electron and the up quark take part in the newly predicted $SU(3)_{grav}$ interaction, whereas the RH sterile neutrino and the RH down quark are singlets of $SU(3)_{grav}$. Moreover, one uses the $SU(3)_{genRH}$ to generate three generations of RH quarks and leptons, and they all have the same square-root mass values as for the first generation: $(0, 1/3, 2/3, 1)$ for sterile neutrino family, electron family, up quark family, down quark family, respectively. Where then do the strange observed mass ratios come from?! Why is the muon 200 times heavier than the electron?

The answer lies in the exceptional Jordan algebra $J_3(8)$. The group $E_6$ happens to be the automorphism group of the complexified exceptional Jordan algebra \cite{Boyle}, and is also the symmetry group of the Dirac equation in ten space-time dimensions \cite{DrayM}. The automorphism group of $J_3(8)$ is $F_4$. We have also argued \cite{Priyank} that $J_3(8)$ is the appropriate algebra for describing three fermion generations.  In the $3\times 3$ matrix with octonionic entries, the diagonal entry (which is a real number) stands for the electric charge [LH fermions], or for square-root mass [RH fermions].  The off-diagonal entries are the octonionic states of the three fermions of a given family.

The characteristic equation of $J_3(8)$, which is a cubic, is of great interest \cite{DrayM}. Its eigenvalues (called Jordan eigenvalues) hold the key to the fundamental parameters of the standard model. For a given value $q$ of the electric charge, the three eigenvalues are $q + \epsilon\sqrt{3/8}$ and for a given value $\sqrt{m}$ of square-root mass, the three eigenvalues are $\sqrt{m}+ \epsilon \sqrt{3/8}$, where $\epsilon = -1, 0, 1$. Since the $U(1)$ charges $q$ and $\sqrt{m}$ arise as a result of symmetry breaking of a unified theory, they must satisfy the condition $q + \sqrt{m}=0, \ \exp[q+\sqrt{m}]=1$. Furthermore, electric charge eigenstates are distinct from eigenstates of square-root mass, and all our measurements, including those of mass, are based on electromagnetism. That is why we observe such strange mass ratios, because square-root mass eigenstates, when expressed in terms of electric charge eigenstates, carry weights which are determined by the aforesaid eigenvalues. These yield the square-root mass ratios as simple but hitherto unanticipated fractions \cite{Singhfsc}, which are distinct from $(0, 1/3, 2/3, 1)$ except for the first generation.

The Jordan eigenvalues, in combination with the Lagrangian of the theory, also yield the limiting low energy value of the fine structure constant $\alpha\equiv e^2/\hbar c$. This is possible because in the Lagrangian $\alpha$ arises as a coefficient resulting from symmetry breaking and separation of the fermionic state into its left-handed and right-handed parts, which are weighted by the Jordan eigenvalues. Consequently, we get that $\alpha = (9/1024) \exp [ 2/3 \times (1/3 - \sqrt{3/8}) ] \approx 1/137.04$. Because of the relation $\exp[q + \sqrt{m}] = 1 $ mentioned above, this implies that the low energy electron has a gravitational fine structure constant $\alpha_g \equiv Gm^2/\hbar c$ given by $\alpha_g^{1/2} = (m/m_{p}) = (9/1024) \exp [ 2/3 \times (\sqrt{3/8} - 1)] \sim 0.010586... = 1/94.4642...$ This gives that the ratio of the strength of the electromagnetic force to the gravitational force is $\alpha / \alpha_g \sim 65.16$. Of course, in the present universe this ratio is $4\times 10^{42}$.

We now explain how cosmological considerations bring  this ratio down to the presently observed value. In order to do so, we borrow only one parameter from observations, this being the total number of particles $N_U$ in the universe, which is known to be about $10^{80}$. The universe has a mass of about $10^{56}$ g. We assume that the universe starts out in an inflationary phase with its initial acceleration equal to the Planck acceleration $a_p \sim 10^{53} \ {\rm cm/s^2}$. Inflation is assumed to end when the universal acceleration, falling inversely as the expansion factor, falls below the surface gravity of a black hole having the same mass as the observed universe, i.e. $10^{56}$ g. Such a black hole has a surface gravity of about $10^{-8}\ {\rm cm/s^2}$ (thus being of the same order as the acceleration $cH_0$ of the present universe, as also the critical MOND acceleration $a_0$). This means that inflation ends after $10^{53} / 10^{-8}=10^{61}$ orders of magnitude scaling up of the expansion factor, taking the universe up in size from $L_p \sim 10^{-33}$ cm to $10^{61}\times 10^{-33} = 10^{28}$ cm as is observed. The condition coming from black hole surface gravity is plausible because it means black holes with size smaller than the universe will have an acceleration higher than the acceleration of the inflating universe, thus allowing classicality to set in, and structures to form.

Our theory obeys the holographic (Karolyhazy) length uncertainty relations: $(\Delta L)^3 \sim L_p^2 \; L$, as a result of which masses scale down by the factor $L^{-1/3}$ with the expansion of the universe. Thus an inflation by $10^{61}$ orders of magnitude brings down the electron's rest mass from its primordial value $m_{p}\alpha_g^{1/2}\sim 0.01 m_p \sim  2\times 10^{-7}$ g by a factor of $10^{61/3}$ to about $0.2 \times 10^{-27}$ g, which is in the correct ballpark range as the observed electron mass. Consequently, the ratio of the strength of the electromagnetic force is brought down by forty-one orders of magnitude, and hence close to the currently observed value. Electric charge values are not scaled down because they belong to the Euclidean sector of Eqn. (\ref{thirteen}) which does not take part in the cosmic expansion. 

Between themselves, the two Lorentzian quadratic forms in Eqn. (\ref{thirteen}) define a 6D spacetime, consistent with $SL(2,\mathbb H)\sim SO(1,5)$, and associated with the interaction $SU(2)_L \times U(1)_Y \times SU(2)_R \times U(1)_{Yg}$. The weak force resulting from the breaking of $SU(2)_L\times U(1)_Y$ is a space-time symmetry associated with the two additional dimensions, whose thickness $L_p \times (L_{univ}/L_p)^{1/3} \sim 10^{-13}$ cm is in the microscopic range, not far from the range of the weak force. This interpretation helps understand why the weak interaction violates parity (a space-time symmetry): only the right-handed particles take part in the $SU(2)_R$ 
pre-gravitational symmetry. Between themselves, the weak force and gravitation together restore 
left-right symmetry. This also helps understand why unlike QCD and electrodynamics, the weak force and gravitation both have dimensional coupling constants, this dimension being inverse mass squared. Moreover, we can expect $SU(2)_R \times U(1)_{Yg}$ to be a renormalizable gauge theory, analogous to the electroweak theory; whereas general relativity just like the weak interaction is not renormalizable, both having resulted from a broken $SU(2)$ symmetry.

From these arguments, we are also able to give a good estimate of the Fermi coupling constant $G^0_F$  of the weak interaction, experimentally measured to be about $10^{-5}$ GeV$^{-2}$ and related to the Higgs vacuum expectation value $v\sim 246$ GeV as $G^0_F \sim 1/v^2$. In this theory, the Higgs  is a composite of the standard model fermions, and its mass is dominantly determined by the heaviest fermions, and hence prior to the onset of inflation it is about $10^4 m_{p} \sim 10^{23}$ GeV. Inflation scales this down by about twenty orders to magnitude to $10^{3}$ GeV and hence resets $v$, and therefore the Fermi constant comes out to be in the range of $10^{-6}$ GeV$^{-2}$. Note that in this theory the gravitational constant $G$, being expressible in fundamental units as $L_p^5 / \tau_p^2 \hbar$ does not change with epoch, and it is the cosmic inflation which renders the weak force so much stronger than gravitation, whereas prior to inflation they both have equal coupling constants, as a part of the unified theory.

This theory clearly predicts that the gravitational sector is $SU(2)_R \times U(1)_{Yg}$ which after symmetry  breaking becomes general relativity modified by a long range $U(1)_{grav}$ interaction. The most important feature of $U(1)_{grav}$ is that its source charge is square root of mass, not mass. The same is true also of the law of MOND acceleration, which is proportional to square-root of the source mass, but for which there is no theoretical reason until we identify MOND with our predicted $U(1)_{grav}$. A striking feature of galaxy rotation curves is that soon as the Keplerian acceleration falls below the critical MOND acceleration $a_0$, the rotation curve turns flat. This feature speaks strongly in favour of MOND, and we are able to reproduce it here from $U(1)_{grav}$. For, upon left-right symmetry breaking, the bosonic dynamical variable $Q_B$ of the theory splits into two parts $Q_B \sim q_B + L \dot{q}_B $, and this is true also for the splitting of $U(1)$ into $U(1)_{em} \sim q_B $ and $U(1)_{grav}\sim L\dot{q}_B$ \cite{Raj}. On dimensional grounds it follows that the potential $\chi$ for $U(1)_{grav}$ changes with distance $R$ as $\ln R$, so that we get the induced acceleration $a = B \sqrt{M} / R$. The constant $B$ is fixed by noting that at the Hubble radius $R_H$ this acceleration must equal the surface gravity $GM_U / R_H^2$ of a black hole (with the same mass as the observed universe), which in turn is equal to $a_0$. Therefore we get at the Hubble radius,
\begin{equation}
a = B \frac{\sqrt{M_U}}{R_H} = \frac{GM_U}{R_H^2} = a_0 \implies B^2 \frac{M_U}{R_H^2} = \frac{GM_U}{R_H^2}\; a_0 \implies B = \sqrt{Ga_0}
\end{equation}
which agrees with the coupling constant in Milgrom's MOND law. We have therefore derived MOND from our newly predicted $U(1)_{grav}$: this interaction dominates over Newtonian gravitation once the Newtonian acceleration falls below $a_0$, forcing the rotation curve to flatten. Moreover, the $U(1)_{grav}$ is based on  a relativistic gauge theory, whose implications for CMB anisotropies are currently under investigation.

In the octonionic theory, the quantum-to-classical transition in the very early universe coincides with the electroweak symmetry breaking, taking place around 1 TeV, which is also when cosmic inflation ends (as discussed above). At this epoch, critical entanglement amongst fermions amplifies the imaginary terms in the space-time part of the line element in Eqn. (\ref{thirteen}), leading to spontaneous localisation and the emergence of space-time.  It is a 6D space-time which contains our 4D observed classical space-time (curved by gravitation: GR + MOND)  as a subset, and the two additional dimensions are microscopically thin and their geometry corresponds to the weak interaction.
Along with this spacetime there emerges the `vector bundle $X$ with a non-abelian group $G$' as in Witten's remark, with the group $G$ being the unbroken symmetry $SU(3)_{color} \times U(1)_{em}$ supplemented by our newly predicted unbroken $SU(3)_{grav} \times U(1)_{grav}$. This $U(1)_{grav}$ provides the theoretical origin of MOND, whereas $SU(3)_{grav}$ is possibly extremely weak and negligible compared to the strong interaction. The $E_8 \times E_8$ unified symmetry has broken into a space-time overlaid by a vector bundle, giving rise also to chiral fermions, and an explanation for the values of (some of the) fundamental constants of the standard model.

Lastly, we note that the 16D split bioctonionic space in Eqn. (3) is possibly directly related to the octonionic projective plane $OP^2$. It is known that there are no octonionic projective geometries for  $n>2$. This strongly suggests that the Lagrangian dynamics that we have constructed on this 16D space is exact, and has no higher order corrections. Beyond the TeV scale we have unification, and this  has been arrived at through a scale-invariant   cosmic inflation beginning at the Planck scale, and ending around 1 TeV \cite{Singh}.

%\bibliography{biblioqmtstorsion}

\end{document}